\documentclass[floatfix,%
 aip,
 amsmath,amssymb,
 reprint,%
]{revtex4-1}

\usepackage{graphicx}% Include figure files
\usepackage{dcolumn}% Align table columns on decimal point
\usepackage{bm}% bold math
\usepackage{comment}
\usepackage[utf8]{inputenc}
\usepackage[T1]{fontenc}
\usepackage{mathptmx}
\usepackage{etoolbox}
\usepackage{xcolor} %GC added
\usepackage{soul} %GC added
\usepackage{hyperref}

\makeatletter
\def\@email#1#2{%
 \endgroup
 \patchcmd{\titleblock@produce}
  {\frontmatter@RRAPformat}
  {\frontmatter@RRAPformat{\produce@RRAP{*#1\href{mailto:#2}{#2}}}\frontmatter@RRAPformat}
  {}{}
}%
\makeatother
\begin{document}

\preprint{AIP/123-QED}

% #!?
\title[]{Magneto-rotation coupling for ferromagnetic nanoelement\\ embedded in elastic substrate}

\begin{comment}
poniżej wkleiłem to co w \textit{Bragg mirror} na arXiv. 
\end{comment}

% #!?
\author{Grzegorz Centa{\l}a}
\email{grzcen@amu.edu.pl}
\author{Jaros{\l}aw W. K{\l}os}
\affiliation{
$^{1}$Institute of Spintronics and Quantum Information, Faculty of Physics and Astronomy,\\ Adam Mickiewicz University, Pozna{\'n},\\
Uniwersytetu~Pozna{\'n}skiego~2, 61-614 Pozna{\'n}, Poland.
}%

\begin{comment}
\author{A. Author}
 \altaffiliation[Also at ]{Physics Department, XYZ University.}%Lines break automatically or can be forced with \\
\author{B. Author}%
 \email{Second.Author@institution.edu.}
\affiliation{ 
Authors' institution and/or address%\\This line break forced with \textbackslash\textbackslash
}%

\author{C. Author}
 \homepage{http://www.Second.institution.edu/~Charlie.Author.}
\affiliation{%
Second institution and/or address%\\This line break forced% with \\
}%
\end{comment}

\date{\today}% It is always \today, today,

\begin{abstract}
This study investigates magneto-rotational coupling as a distinct contribution to magnetoelastic interactions, which can be influenced by magnetic anisotropy. We determine magneto-rotational coupling coefficients that incorporate the shape anisotropy of a magnetic nanoelement (strip) and demonstrate that this type of coupling can be modified through geometric adjustments. Furthermore, we analyze the magneto-rotational contribution to the magnetoelastic field in a ferromagnetic strip embedded in a nonmagnetic substrate. Both Rayleigh and Love waves are considered sources of the magnetoelastic field, and we examine how the strength of the magneto-rotational coupling varies with the direction of the magnetization, and the aspect ratio of the strip cross-section. We analyze the changes of the magneto-rotational contribution to the magnetoelastic field with an increasing thickness-to-width ratio, assuming a fixed magnetization direction corresponding to the strongest magnetoelastic coupling. For Love wave, the contribution of the out-of-plane component increases monotonically, while that of the in-plane component decreases monotonically. In the case of the Rayleigh wave, only the out-of-plane component contributes, and it approaches zero as the cross-section becomes square. These findings enhance the understanding of magneto-rotational coupling in magnonic nanostructures.
\end{abstract}

\maketitle

%\begin{quotation}
%The ``lead paragraph'' is encapsulated with the \LaTeX\ 
%\verb+quotation+ environment and is formatted as a single paragraph before the first section heading. 
%(The \verb+quotation+ environment reverts to its usual meaning after the first sectioning command.) 
%Note that numbered references are allowed in the lead paragraph.
%
%The lead paragraph will only be found in an article being prepared for the journal %\textit{Chaos}.
%\end{quotation}

\section{\label{sec:level1}Introduction}
Based on the wave computing paradigm\cite{zangeneh-nejad_analogue_2021,Mahmoud_2020}, spin waves and surface acoustic waves (SAWs) enable the design of nanoscale magnonic\cite{Chumak_2022} and phononic\cite{campbell_surface_1998,Delsing_2019} devices that process GHz signals. This approach allows for the implementation of computational schemes that are difficult or even impossible to achieve with conventional electronic circuits, such as neuromorphic computing\cite{papp_nanoscale_2021,Kraimia_2020} or even the efficient simulation of quantum algorithms\cite{Yang_2021}. However, wave computing on both magnonic and phononic platforms faces unique challenges. For phononics, achieving the nonlinear regime or non-reciprocal propagation is a significant hurdle, while for magnonics, relatively low group velocity and high attenuation present notable limitations. Hybrid magnonic-phononic systems offer a promising solution to overcome these challenges.

Typically, magnonic-phononic hybrids\cite{Thevenard_2016,Verba_2018,Xu_2018,Mondal_2018,Latcham_2019,Kuss_2020,Tateno_2020,Yokouchi_2020,Babu_2021,Seemann_2022,Geilen2022,Lopes_2024,seeger2024} rely on magnetic materials with strong magnetostrictive properties due to their microscopic (atomic) structure. This requirement limits the choice of magnetic materials, as they must also exhibit relatively low damping of magnetization dynamics. An intriguing alternative involves using magnetic materials and structures characterized by magnetocrystalline or shape anisotropy, which enable the exploitation of magneto-rotational coupling\cite{Xu_2020,Sato_2021}. This unconventional magnetoelastic interaction not only facilitates coupling but also induces a non-reciprocity effect\cite{Sato_2021,Liao_2024}.

The magneto-rotational coupling has been a well-known phenomenon for nearly 50 years\cite{Maekawa_1976,Baryakhtar_1985}, with its theoretical foundations established in the 1960s\cite{mindlin_effects_1962,Tiersten_1965}. Recently, this narrow field has experienced a revival in both experimental\cite{Xu_2020,Liao_2024} and theoretical research\cite{Sato_2021, Yammamoto_2022}, driven by increasing interest in magnetoelastic systems that explore the interplay between SAWs and spin waves in magnetic layers, as initiated by Weiler\cite{Weiler_2011,Dreher_2012}.

Most prior work on magneto-rotational coupling focuses on homogeneous magnetic layers with magnetocrystalline anisotropy deposited on non-magnetic substrates\cite{Sato_2021}. In such cases, shape anisotropy is determined solely by the saturation magnetization and the orientation of magnetization relative to the surfaces. In contrast, the work presented here investigates magneto-rotational coupling between the fundamental mode of precessing magnetization and SAWs in a ferromagnetic strip embedded in an elastic, non-magnetic substrate. Specifically, we examine how varying the strip's shape (defined by the ratio of its thickness to width) affects magneto-rotational coupling with Rayleigh and Love waves. Our research shows that it is possible to modify the magneto-rotational coupling by changing the shape anisotropy of the ferromagnetic nanoelement. We found that the magneto-rotational contribution to the magnetoelastic field changes differently with the thickness-to-width ratio for Rayleigh and Love waves.

In the Model section, we introduce the formalism used to determine the magneto-rotational coupling coefficients for the strip, considering the dynamic magnetoelastic contributions and the magneto-rotational effect. In the Results section, we present and analyze the dependence of magneto-rotational contributions to magnetoelastic energy and fields on the orientation angle of the equilibrium magnetization.

\section{The Model}
The magneto-rotation coupling is related to the presence of magnetic anisotropy in magnetic material which experiences elastic deformation in the form of local twists. Such deformation is formally described by the non-zero antisymmetric part $\mathbf{\omega}=\tfrac{1}{2}(\nabla \mathbf{u}-\nabla \mathbf{u}^{T})$ of displacement gradient tensor $\nabla \mathbf{u}$ and, in general approach, gives the contribution to magnetoelastic energy density $G_{\rm me}$. The rotation tensor $\mathbf{\omega}$ is often neglected because the equilibrium condition for the whole body requires the balance of the mechanical torques. However, the precessing magnetization can be a source of the torque\cite{Tiersten_1965} and $\mathbf{\omega}$ can not be omitted for magnetoelastic systems.

The magnetoelastic energy density in continuum and elastically isotropic medium, expended up to linear terms in strain  $\mathbf{\varepsilon}=\tfrac{1}{2}(\nabla \mathbf{u}+\nabla \mathbf{u}^{T})$ and rotation tensors $\mathbf{\omega}$, is given by the formula:
\begin{equation}
    G_{\rm me}=\sum_{\alpha,\beta}\left(b_{\alpha\beta}\varepsilon_{\alpha\beta}+K_{\alpha\beta}\omega_{\alpha\beta}\right)m_{\alpha}m_{\beta}\label{eq:magel_ener},
\end{equation}
where the coefficient $b_{\alpha\beta}$ describes the conventional magnetoelastic interaction and $K_{\alpha\beta}$ magneto-rotation coupling. Since $G_{\rm me}$ is a quadratic from $\mathbf{m}^{\rm T}\!\cdot\!\mathbf{A}\cdot\!\mathbf{m}$ in terms of magnetization vector $\mathbf{m}=\mathbf{M}/M_{\rm s}$ (normalized to saturation magnetization $M_{\rm s}$), the matrix $\mathbf{A}$ can be uniquely defined as symmetric matrix: $A_{\alpha\beta}=b_{\alpha\beta}\varepsilon_{\alpha\beta}+K_{\alpha\beta}\omega_{\alpha\beta}=A_{\beta\alpha}$. Taking into account that the matrix of strain (and rotation) is symmetric (antisymmetric) by definition: $\varepsilon_{\alpha\beta}=(\partial_\beta u_\alpha+\partial_\alpha u_\beta)/2=\varepsilon_{\beta\alpha}$, ($\omega_{\alpha\beta}=(\partial_\beta u_\alpha-\partial_\alpha u_\beta)/2=-\omega_{\beta\alpha}$), we can find that corresponding matrices of coefficients must be symmetric $b_{\alpha\beta}=b_{\beta\alpha}$ (and antisymmetric $K_{\alpha\beta}=-K_{\alpha\beta}$).

 The magneto-rotation coupling results from the fact that the anisotropy axis in magnetic material $\hat{\mathbf{n}}$ changes its direction $\hat{\mathbf{n}}\rightarrow\hat{\mathbf{n}}+\delta\hat{\mathbf{n}}$ due to elastic deformation, i.e. rotates around the axis of the elastic twist by the angle  $\delta\varphi=\tfrac{1}{2}\nabla\times\mathbf{u}$, which modifies $\hat{\mathbf{n}}$ by the amount   $\delta \mathbf{n}=\delta\mathbf{\varphi}\times\hat{\mathbf{n}}$. The angle $\delta\varphi(\mathbf{\omega})$ depends on the components $\omega_{\alpha\beta}$ of the rotation tensor. Therefore, such correction to anisotropy energy density can be interpreted as a contribution $K_{\alpha\beta}\omega_{\alpha\beta}m_{\alpha}m_{\beta}$ to magnetoelastic energy density.

The magnetic anisotropy has two main sources: (i) volume and surface magnetocrystalline anisotropy, related to the atomistic ordering of the magnetic material, and (ii) shape anisotropy, generated by demagnetizing effects of the magnetic body. Regardless of the source magnetic anisotropy, it will generate magneto-rotation coupling which can be incorporated in the general equation (\ref{eq:magel_ener}). These properties refer to any component of the energy density that is characterized by anisotropic dependence on $\mathbf{m}$.

\begin{figure}
\includegraphics[width=0.4\textwidth]{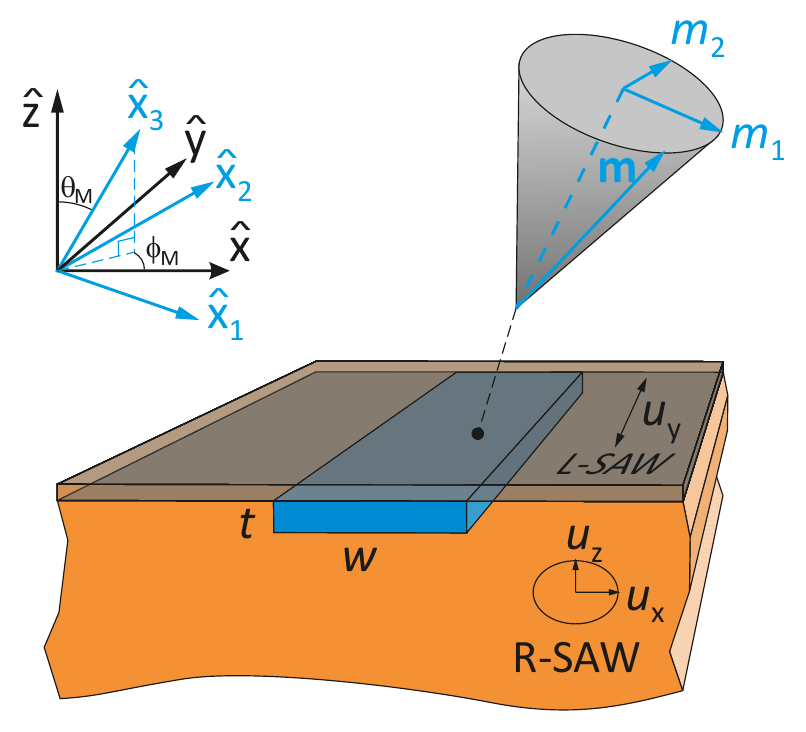}% Here is how to import EPS art
\caption{\label{fig:epsart} Magnetoelastic interaction between the fundamental mode of the precessing magnetization in a ferromagnetic strip (blue) and surface acoustic waves (SAW) propagating in a non-magnetic substrate (orange) along the $x$-direction, i.e. perpendicular to the strip.  The interaction is not only due to the intrinsic magnetostriction of the ferromagnetic material but also caused by the magnetic anisotropy and related to the magneto-rotation coupling. The magnetic anisotropy can be tuned by modifying the shape anisotropy, determined by the ratio of thickness $t$ to width $w$, and the magnetocrystalline anisotropy, introduced by interfacing the ferromagnet with another material (gray layer). The magnetoelastic interaction is strongly anisotropic and depends on both the direction of the equilibrium magnetization $\hat{\mathbf{x}}_3$ (i. e. the orientation of the precession axis is given by $\theta_{M}$ and $\phi_{M}$ angles), controlled by the external magnetic field, and the polarization of the SAW -- the interaction is different for Love-SAW (L-SAW) and Rayleigh-SAW (R-SAW). }\label{fig:structure}
\end{figure}

In our studies, we considered the simple ferromagnetic nanoelement (strip of the width $w$ and thickness $t$) deposited on a non-magnetic substrate where the surface acoustic waves (SAW) can propagate with in-plane applied magnetic field -- see Fig.~\ref{fig:structure}. The strip is characterized by both the shape anisotropy, tending to align the magnetization along the strip, and surface out-of-plane anisotropy $K_{\rm s}$, on its bottom or top face. The effective magnetocrystalline anisotropy $K_{1}=K_{\rm s}/t$ depends on the thickness and the shape anisotropy on the thickness-to-width ratio $p=t/w$. As a result, the magneto-rotation coupling is quite complex and can be tuned by geometric means. We considered the coupling between SAW and the fundamental mode of precessing magnetization. The density of energy related to anisotropy  can be written in the general form:
\begin{equation}
    G_{\rm a}=\tfrac{1}{2}\mu_0 M_{\rm s}^2 \;\mathbf{m}^{\rm T}\!\!\cdot\mathbf{N}\cdot\!\mathbf{m}+K_1(\mathbf{m}\times\hat{\mathbf{n}}_{\rm mc})^2,\label{eq:anizoen_gen}
\end{equation}
where $\mu_0$ is vacuum permeability and the magnetocrystalline uniaxial anisotropy is oriented along $z-$axis: $\hat{\mathbf{n}}_{\rm mc}=\hat{\mathbf{z}}$. 

In $xyz$-Cartesian coordinate system, the demagnetizing tensor $\mathbf{N}$ for the strip, oriented as presented on Fig.~\ref{fig:structure} has only two non-zero elements\cite{Aharoni_1998}:
\begin{equation}
\begin{split}
     N_{xx}&=\frac{1}{\pi}\!\left(\frac{p\!-\!p^{-1}}{2}\ln(1+p^{-2})+p^{-1}\ln(p^{-1})+2\arctan(p)\right),\\
     N_{zz}&=1-N_{xx}.
\end{split}\label{eq:N}
\end{equation}

It is worth noting that the demagnetization energy density formula $\tfrac{1}{2}\mu_0\, M_{\rm s}^2\mathbf{m}^{\rm T}\!\!\cdot\mathbf{N}\cdot\!\mathbf{m}$ is strict for a generalized ellipsoid\cite{Osborn_1945}, e.g. for an elliptical strip. However, this is a very good approximation for square strip ($p\approx 1$), which also holds for flatter strips (e.g. for the structure considered here where $p\approx 0.1$), if we can still neglect the dipolar pinning\cite{Centala2019,rychly_2022,Centala2023}.
%For the case of the elliptical strip, we have to recalculate the effective uniaxial anisotropy and use a different formula for the components of the demagnetization tensor: $N_{yy}=p/(1-p)$, $N_{zz}=1-N_{yy}$, where $p$ is a ratio of short to long semi-axis\cite{Osborn_1945}.

If the shape anisotropy dominates over the magnetocrystalline anisotropy ($ \tfrac{1}{2}\mu_0 M_{\rm s}^2 N_{zz} - K_1 > 0$), then uniaxial easy-plane anisotropy can be introduced for both the $x$-- and $z$--directions. Next, we consider how the rotations of the versors, $\hat{\mathbf{x}} \rightarrow \hat{\mathbf{x}} + \delta\hat{\mathbf{x}}$ and $\hat{\mathbf{z}} \rightarrow \hat{\mathbf{z}} + \delta\hat{\mathbf{z}}$, modify the energy density associated with magnetic anisotropy. Taking into account (\ref{eq:N}), we can express the anisotropy energy density (\ref{eq:anizoen_gen}) in the presence of an elastic twist of the magnetic material as:
\begin{equation}
\begin{split}
     G_{\rm a}=K_1&+\tfrac{1}{2}\mu_0 M_{\rm s}^2 N_{xx}\,\big(\mathbf{m}\cdot(\hat{\mathbf{x}}+\delta\hat{\mathbf{x}})\big)^2\\
     &+\left( \tfrac{1}{2}\mu_0 M_{\rm s}^2 N_{zz}-K_1\right)\big(\mathbf{m}\cdot(\hat{\mathbf{z}}+\delta\hat{\mathbf{z}})\big)^2.\label{eq:anizoen_2}
\end{split}
\end{equation}
Taking advantages from the fact that changes of the directions $\delta\hat{\mathbf{x}}$, and $\delta\hat{\mathbf{z}}$ of the $xyz$-axis are small, we can write (\ref{eq:anizoen_2}):
\begin{equation}
\begin{split}
     G_{\rm a}&\!=K_1\!+\!\tfrac{1}{2}\mu_0 M_{\rm s}^2 N_{xx}\,(\mathbf{m}\cdot\hat{\mathbf{x}})^2
     \!+\!\left( \tfrac{1}{2}\mu_0 M_{\rm s}^2 N_{zz}\!-\!K_1\right)\!(\mathbf{m}\cdot\hat{\mathbf{z}})^2\\
     &+\mu_0 M_{\rm s}^2 N_{xx}(\omega_{yx}m_{y}m_{x}+\omega_{zx}m_{z}m_{x})\\
     &+\left(\mu_0 M_{\rm s}^2 N_{zz}-2K_1\right)(\omega_{xz}m_xm_z+\omega_{yz}m_ym_z).
     \label{eq:anizoen_3}
\end{split}
\end{equation}

The last two terms in (\ref{eq:anizoen_3}) is the magneto-rotational contribution to magnetoelastic energy density $G_{\rm me}$. The first three terms denote the anisotropy energy density in the absence of deformation -- compare to (\ref{eq:anizoen_2}) for $\delta\hat{\mathbf{x}}=0$ and $\delta\hat{\mathbf{z}}=0$. When deriving of (\ref{eq:anizoen_3}), we assumed that  $\delta\hat{\mathbf{x}}=\omega_{yx}\hat{\mathbf{y}}+\omega_{zx}\hat{\mathbf{z}}$ and $\delta\hat{\mathbf{z}}=\omega_{xz}\hat{\mathbf{x}}+\omega_{yz}\hat{\mathbf{y}}$ are small and neglected the terms quadratic in  $\delta\hat{\mathbf{x}}$ and $\delta\hat{\mathbf{z}}$. 

\begin{figure}[t]
\includegraphics[width=0.4\textwidth]{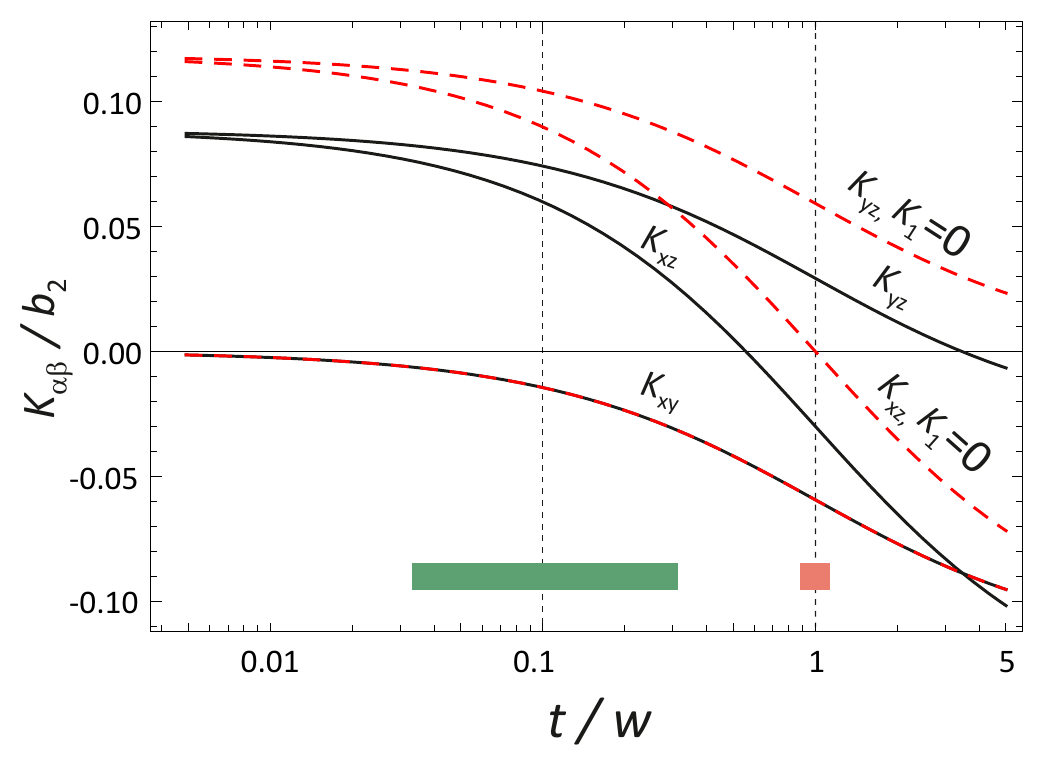}% Here is how to import EPS art
\caption{\label{fig:epsart} The coefficients $K_{\alpha\beta}$ for the magneto-rotation coupling as a function of the aspect ratio thickness/width ($t/w$) of the ferromagnetic strip. The values of $K_{\alpha\beta}$ obtained from the equations (\ref{eq:K}) and (\ref{eq:N}) are related to the conventional magneto-elastic constant $b_2=7$~MJ/m$^3$. We have fixed the thickness of the strip $t=5$~nm and varied its width $w$. The solid black (dashed red) lines denote the case where the magnetocrystalline anisotropy $K_1=K_{\rm s}/t$ is present (absent). The calculation was performed for surface anisotropy $K_{\rm s}=1.05$~mJ/m$^2$ and saturation magnetization $M_{\rm s}=1150$~kA/m. The green rectangle and pink square visualize the aspect ratio. In the absence of $K_1$, the absolute sizes $t$, $w$ are irrelevant and the coefficients $K_{\alpha\beta}$ are determined only by the aspect ratio $t/w$.}\label{fig:K}
\end{figure}

By comparing the magneto-rotational contribution in (\ref{eq:anizoen_3}) to its general form $K_{\alpha,\beta}\omega_{\alpha,\beta}m_\alpha m_\beta$ in (\ref{eq:magel_ener}), we can determine the coefficients $K_{\alpha\beta}$ for magneto-rotation coupling:
\begin{equation}
\begin{split}
K_{xy}&=-\tfrac{1}{2}\mu_0 M_{\rm s}^2 N_{xx}=-K_{yx}\\
K_{xz}&=\tfrac{1}{2}\mu_0 M_{\rm s}^2 (N_{zz}-N_{xx})-K_1=-K_{zx}\\
K_{yz}&=\tfrac{1}{2}\mu_0 M_{\rm s}^2 N_{zz}-K_1=-K_{zy}\\
K_{xx}&=K_{yy}=K_{zz}=0.
\end{split}\label{eq:K}
\end{equation}
The coefficients, $K_{\alpha,\beta}$, depend on the geometry of the ferromagnetic nanoelement because they are expressed in terms of the demagnetizing tensor, as shown in Fig.~\ref{fig:K}.

The conventional magnetoleastic coupling constants, resulting from isotropic magnetostriction of magnetic material are given by the formula: $b_{\alpha\beta}=\delta_{\alpha\beta}b_1+(1-\delta_{\alpha\beta})b_2$, where $\delta_{\alpha\beta}$ is Kronecker delta.

Let's discuss now the magnetoelastic energy density (\ref{eq:anizoen_gen}) for dynamic magnetization precessing around the arbitral direction $\hat{\mathbf{x}}_3$, determined by the anisotropy and applied field. We assume that the equilibrium direction for static magnetization $\hat{\mathbf{x}}_3$ is deflected from $\hat{\mathbf{z}}$-direction by the angle $\theta_{M}$, and its projection on $xy$-plane creates the angle $\phi_{M}$ with $\hat{\mathbf{x}}$-direction -- see also Appendix~\ref{Appendix A}. Then, we can consider the magnetization vector $\mathbf{m}'$ in $x_1x_2x_3$ Cartesian coordinate system rotated by the angles $\theta_{M}, \phi_{M}$ respect to $xyz$ system -- see Fig.~\ref{fig:structure}. In the linear approximation, the component of magnetization along the equilibrium direction can be considered as constant and equal to saturation magnetization $m_3\approx1$ and the remaining dynamic components are small:   $m_1(t),m_2(t)\ll 1$. The transformation of magnetization vector between $xyz$ and $x_1x_2x_3$ coordinates systems: $\mathbf{m}=\mathbf{R}\cdot\mathbf{m}'$ is given by the orthonormal matrix $\mathbf{R}^{-1}=\mathbf{R}^{\rm T}$:
\begin{equation}
\mathbf{R}=
    \begin{pmatrix}
\cos\theta_{M}\cos\phi_{M} & \sin\phi_{M} & \sin\theta_{M}\cos\phi_{M}\\
\cos\theta_{M}\sin\phi_{M} & \cos\phi_{M} & \sin\theta_{M}\sin\phi_{M}\\
-\sin\theta_{M} & 0 & \cos\theta_{M}
\end{pmatrix}.
\end{equation}
The transformation of the matrix $A_{\alpha\beta}=b_{\alpha\beta}\varepsilon_{\alpha\beta}+K_{\alpha\beta}\omega_{\alpha\beta}$ from $xyx$ to $x_1x_2x_3$ coordinate system is expressed as: $\mathbf{A}'=\mathbf{R}^{-1}\cdot\!\mathbf{A}\cdot\mathbf{R}$. This allows finding the leading term of the magnetoelastic energy density ${g_{\rm me}}$ depending on dynamic components of magnetization $m'_1$, $m'_2$:
\begin{equation}
\begin{split}
    G_{\rm me}&=\mathbf{m}^{\rm T}\!\cdot\!\mathbf{A}\cdot\!\mathbf{m}=\mathbf{m}'\,^{\rm T}\!\cdot\!\mathbf{A}'\cdot\!\mathbf{m}'\\
    &=A'_{33}\\&\underbrace{+2A'_{13}m'_{1}+2A'_{23}m'_2}_{g_{\rm me}}\\&+2A'_{12}m'_1m'_2+A'_{11}m'^2_1+A'_{22}m'^2_2,
\end{split}
\end{equation}
where we took $\mathbf{m}'=m_1\hat{\mathbf{x}}_1+m_2\hat{\mathbf{x}}_2+\hat{\mathbf{x}}_3$ and used the identity: $(\mathbf{R}\cdot\mathbf{m}')^{\rm T}=\mathbf{m}'^{\,\rm T}\cdot\mathbf{R}^{\rm T}$. The  expression for ${g_{\rm me}}$ reads:
\begin{equation}
    g_{\rm me}=2(A_{13}' m'_1+A_{23}' m'_2), \label{eq:anizoen_4}
\end{equation}
where $A_{13}'$ and $A_{23}'$ takes the explicit form:
\begin{equation}
\begin{split}
A_{13}'=\tfrac{1}{4} \sin (2 \theta_{M} ) \big(&b_1 (\varepsilon_{xx}+\varepsilon_{yy}-2 \varepsilon_{zz})\\
 +&b_1 ( \cos (2 \phi_{M} )\varepsilon_{xx}-\varepsilon_{yy})\\
 +&2 \sin (2\phi_{M} ) (b_2 \varepsilon_{xy}+K_{xy} \omega_{xy})\big)\\
 +\cos (2 \theta_{M} ) \big(&\cos (\phi_{M} ) (b_2 \varepsilon_{xz}+K_{xz} \omega_{xz})\\
 +&\sin (\phi_{M} )(b_2 \varepsilon_{yz}+K_{yz} \omega_{yz})\big),\\
 A_{23}'=\sin (\theta_{M} ) \big(&\tfrac{1}{2}\sin (2\phi_{M} ) b_1 (\varepsilon_{yy}-\varepsilon_{xx}) \\
+&\cos (2 \phi_{M} ) (b_2 \varepsilon_{xy}+K_{xy} \omega_{xy})\big)\\
+\cos(\theta_{M} ) \big(&\cos (\phi_{M} ) (b_2 \varepsilon_{yz}+K_{yz} \omega_{yz})\\
-&\sin (\phi_{M} ) (b_2 \varepsilon_{xz}+K_{xz} \omega_{xz})\big).
\end{split} \label{eq:anizoen_coeff}
\end{equation}
The magnetoelastic energy density $G_{\rm me}$ can be used to determine the  contribution to effective field perceived by magnetization as a result of magnetoelastic coupling:
\begin{equation}
    \mathbf{H}_{\rm me}=-\frac{1}{\mu_0 M_{\rm s}}\nabla_\mathbf{m}G_{\rm me},
\end{equation}
where $\nabla_\mathbf{m}$ is the gradient taken respect to the components of $\mathbf{m}$. The magnetoelastic field $\mathbf{H}_{\rm me}$ is introduced to the linearized Landau-Lifshitz equation as an external field which does not depend on magnetization and is determined by the gradient of dynamic deformation: $\mathbf{\varepsilon}$, $\mathbf{\omega}$. We should calculate the magnetoelastic field in $x_1x_2x_3$ coordinate system at the equilibrium orientation of magnetization 
 $\mathbf{m}'_0=\hat{\mathbf{x}}_3$.
\begin{equation}
    \mathbf{H}'_{\rm me}=-\frac{1}{\mu_0 M_{\rm s}}\nabla_\mathbf{m'}G_{\rm me}\big|_{\mathbf{m}'=\mathbf{m}'_0}=-\frac{2}{\mu_0 M_{\rm s}}\mathbf{A}'\cdot\mathbf{m}'\big|_{\mathbf{m}'=\mathbf{m}'_0},
\end{equation}
where we used the following identity for quadratic from defined by symmetric matrix: $\nabla_\mathbf{m'}(\mathbf{m}'^{\rm\; T}\cdot\mathbf{A}'\cdot\mathbf{m}')=2\,\mathbf{A}'\cdot\mathbf{m}'$. The components of $\mathbf{H}'_{\rm me}$ taken in $x_1$- and $x_2$-directions read:
\begin{eqnarray}
    h'_{1,\rm me}=-\frac{2}{\mu_0 M_{\rm s}}A'_{13}, &&  h'_{2,\rm me}=-\frac{2}{\mu_0 M_{\rm s}}A'_{23}.\label{eq:mefield}
\end{eqnarray}

\section{Results}

We considered the ferromagnetic CoFeB strip, where the surface anisotropy was induced by the MgO layer covering the strip embedded in an elastic substrate -- see Fig.~\ref{fig:structure}. For such a system, we took the following values of material parameters: surface anisotropy:\cite{Lee_2014} $K_{\rm s}=1.05$~mJ/m$^2$ and saturation magnetisation:\cite{Lee_2014} $M_{\rm s}=1150$~kA/m. We assumed the magnetoelastic coupling constants\cite{vanderveken2022}$^,$\footnote{According to [\onlinecite{Kuss_2020}], the values of $b_1$, expressed in teslas, range from -3.8 to -6.5 T for CoFeB. Dividing this values by $\mu_0 M_{\rm s}$, we obtain the corresponding values in J/m$^3$: 5.5 -- 9.4 MJ/m$^3$}: $b_1=b_2=7$~MJ/m$^3$.

\begin{figure}[!h]
\includegraphics[width=0.51\textwidth]{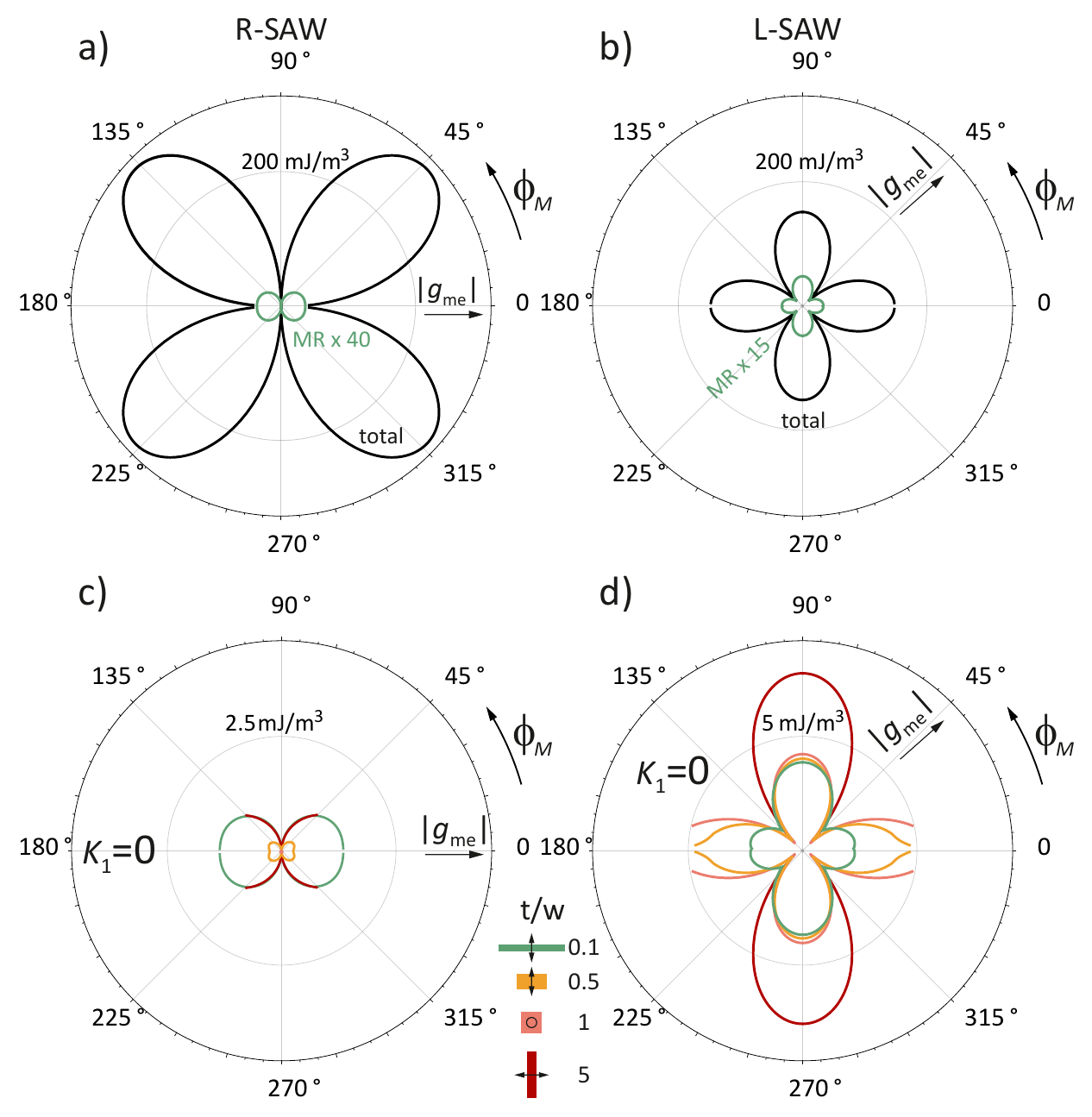}% Here is how to import EPS art
\caption{The angular dependence of the magnetoelastic energy density $|g_{\rm me}|$ (see Eqs.~\ref{eq:anizoen_4} and \ref{eq:anizoen_coeff}) for the case when the equilibrium magnetization is oriented in the plane ($\theta_{M}=90^{\rm o}$).  We consider the coupling of the magnetization dynamics with surface acoustic waves of different polarization (a,c) R-SAW and (b,d) L-SAW -- see Fig.~\ref{fig:structure}. In (a,b), we present the density of magnetoelastic energy for the ferromagnetic strip of width $w=50$~nm and thickness $t=5$~nm, using the same material parameters as in Fig.~\ref{fig:K}.  Black (green) lines represent the total value of $|g_{\rm me}|$ (the contribution of the magneto-rotation coupling to $|g_{\rm me}|$, corresponding to the case when the ferromagnet has no intrinsic magnetostriction $b_1=b_2=0$). The contribution of the magneto-rotation coupling is small, and therefore all green contours are magnified 40 times (a) or 15 times (b). For (c,d), we neglected the magnetocrystalline anisotropy $K_1=0$ and we change the shape anisotropy. Color lines show the contribution of the magneto-rotation coupling for different values of the thickness-to-width ratio ($t/w$), arrows (circle) indicate the hard anisotropy axis (lack of hard anisotropy axis) -- see inset between (c) and (d). It should be noted that different energy scales are used in (c) and (d). The dynamic magnetization amplitudes were calculated numerically (see Appendix B); we took the following averaged values of the elements of the strain and rotation tensors: $\varepsilon_{xx}=\varepsilon_{yy}=10^{-6}$, $\varepsilon_{zz}=0.1\varepsilon_{xx}$, $\varepsilon_{xy} = 0. 25\varepsilon_{xx}$, $\varepsilon_{xz}=\varepsilon_{yz}$ $= 0.05 \varepsilon_{xx}$, $\omega_{xy}=\varepsilon_{xy}$, $\omega_{xz}=\omega_{yz}=\varepsilon_{xz}$.}\label{fig:g}
\end{figure}

The magnetoelastic interaction is characterized by a strong anisotropy. It depends both on the direction around which the magnetization precesses and on the polarization of the elastic waves. It seems interesting to estimate the influence of the magneto-rotation interaction on this anisotropy or to determine it qualitatively in the absence of conventional magnetoelasticity $b_1=b_2=0$. For the assumed values of $ M_{\rm s}$, $K_{\rm s}$ and $t$ the shape anisotropy prevails over the magnetocrystalline anisotropy (\ref{eq:anizoen_gen}), which means that when an external magnetic field is applied in the plane of the strip, the equilibrium magnetization remains oriented in the plane ($\theta_{M}=90^{\rm o}$), between the strip axis and the field direction. This makes it possible to simplify the study and to consider the anisotropy of the magnetoelastic (and especially magneto-rotation) interaction as a function of the direction $\phi_{M}$ of the plane-oriented equilibrium magnetization of the strip. For this geometry, the dynamic components of the magnetization $m'_1$ and $m'_2$ are oriented in the out-of-plane and in-plane directions, respectively, which means that they will differ in amplitude. This ratio varies with the orientation of the equilibrium magnetization and the applied field H$_{0}$ -- see Appendix~\ref{Appendix B}. The dynamic magnetization amplitudes were calculated numerically. In Fig.~\ref{fig:g}, we plot the angular dependence of the magnetoelastic interaction energy density estimated as $|g_{\rm me}|\approx |A'_{13}m'_1|+ |A'_{23}m'_2|$ where the following averaged values of the strain and rotational tensor elements were taken (we assume at the wavelength of SAW is larger than the strip width):  $\varepsilon_{xx}=\varepsilon_{yy}=10^{-6}$, $\varepsilon_{zz}=0.1\varepsilon_{xx}$, $\varepsilon_{xy} = 0. 25\varepsilon_{xx}$, $\varepsilon_{xz}=\varepsilon_{yz}$ $= 0.05 \varepsilon_{xx}$, $\omega_{xy}=\varepsilon_{xy}$, $\omega_{xz}=\omega_{yz}=\varepsilon_{xz}$\cite{Dreher_2012}. For the calculations of the energy density of the magnetoelastic interaction, we have considered the very small amplitude of the SW precession obtained from numerical solutions of the linearised Landau-Lifshtz equation -- see Appendix B. The values of $\mathbf{m}=\mathbf{M}/M_{\rm s}$ are of the order of $10^{-3}$.

We considered two particular polarizations of the SAW Love-SAW (L-SAW) and Rayleigh-SAW (R-SAW). For considered geometry (Fig.~\ref{fig:structure}) the following elements of strain (and rotation) tensors are non-zero for (i) R-SAW: $\varepsilon_{xx}$, $\varepsilon_{zz}$, $\varepsilon_{zx}=\varepsilon_{xz}$ ($\omega_{zx}=-\omega_{xz}$), 
 and (ii) L-SAW: $\varepsilon_{xy}=\varepsilon_{yx}$ ($\omega_{xy}=-\omega_{yx}$), $\varepsilon_{yz}=\varepsilon_{zy}$ ($\omega_{yz}=-\omega_{zy}$).

 In Fig.~\ref{fig:g}, we presented the angular dependence of the elastic energy density $|g_{\rm me}|(\phi_{M})$ for different in-plane ($\theta_{M}=\pi/2$) orientation $\phi_{M}$ of equilibrium magnetization $M_{\rm s}\hat{\mathbf{x}}_3$ ($\phi_{M}=$90$^{\circ}$ means that the $\hat{\mathbf{x}}_3$ is oriented along the CoFeB strip $\hat{\mathbf{x}}_3=\hat{\mathbf{y}}$). In Fig.~\ref{fig:g}(a,b), we presented the case of the flat strip ($t/w=0.1$) with out-of-plane magnetocrystalline anisotropy $K_1$ which competes with its shape anisotropy. The green contours present a small contribution from magneto-rotation coupling which is magnified 40 times (Fig.~\ref{fig:g}(a)) or 15 times (Fig.~\ref{fig:g}(b)) in reference to total magnetoelastic energy density (black contour). We can see that for R-SAW, the magneto-rotation coupling enhances the total magnetoelastic energy density $|g_{\rm me}|(\phi_{M})$ and shallows its minimum at the direction $\phi_{M}=0$,  where the equilibrium magnetization is perpendicular to the strip's axis. On the other hand, for L-SAW, the magneto-rotation increases $|g_{\rm me}|(\phi_{M})$ by a few percent in the direction $\phi_{M}=90^{\rm o}$ where $|g_{\rm me}|$ was already maximized. It is worth noting that this direction ($\phi_{M}=90^{\rm o}$) is the easy axis of shape anisotropy of the strip and we do not need to apply external magnetic field $H_0$ to align the equilibrium magnetization along this direction.

  \begin{figure}
\includegraphics[width=0.52\textwidth]{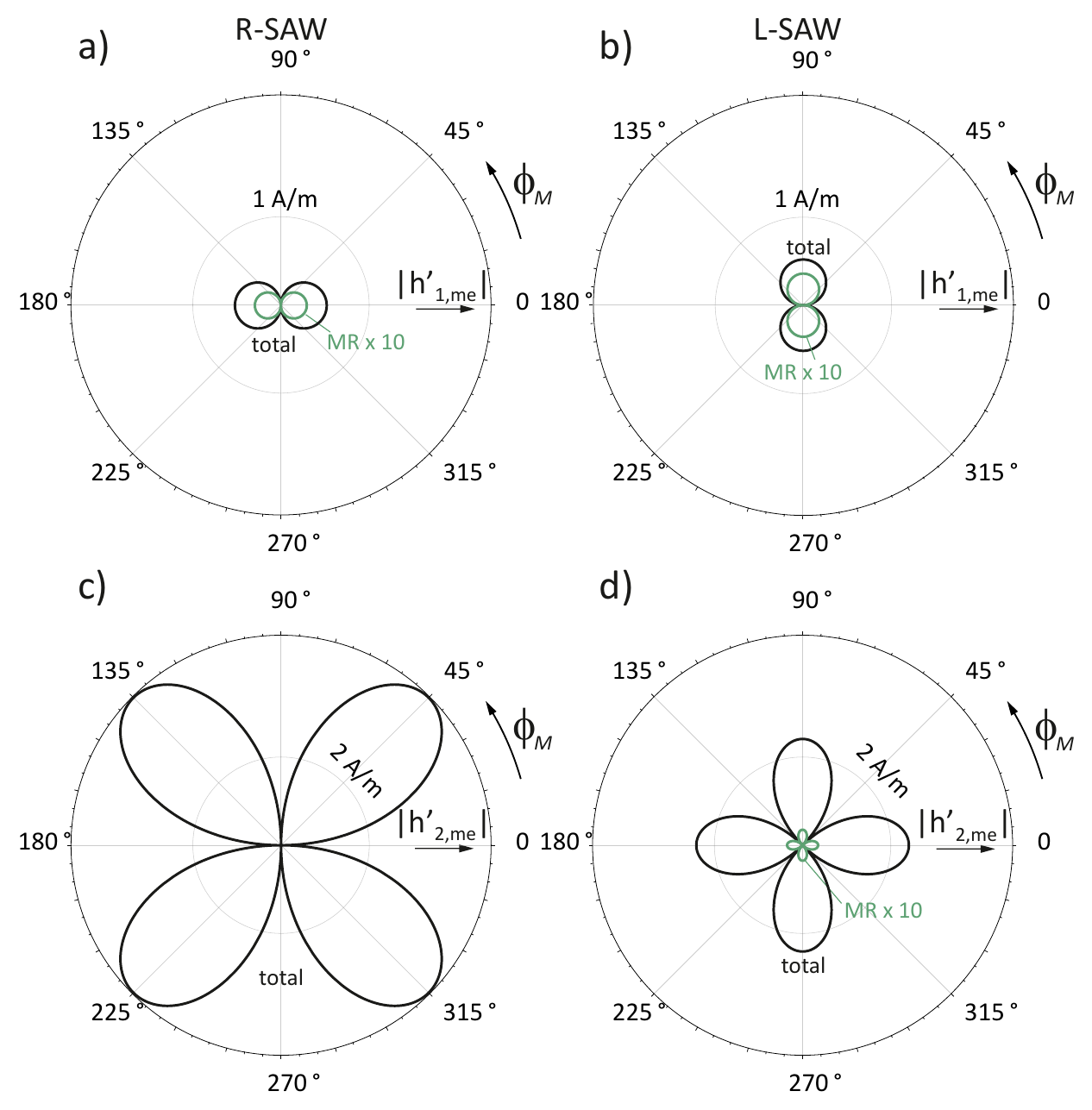}% Here is how to import EPS art
\caption{The angular dependence of the dynamic components of the magnetoelastic field $|h'_{i,\rm me}|$ (\ref{eq:mefield}) for the case when the equilibrium magnetization is oriented in the plane: ($\theta_{M}=90^{o}$). We consider the coupling of the magnetization dynamics with surface acoustic waves of different polarization (a,c) R-SAW and (b,d) L-SAW -- see Fig.~\ref{fig:structure}. The upper (a,b) and lower (c,d) panels show the results for the out-of-plane component $|h_{1,{\rm me}}|$ and the in-plane component $|h_{2,{\rm me}}|$, respectively. The black (green) lines represent the total values of $|h'_{i,\rm me}|$ (the contribution of the magneto-rotation coupling $|h'_{i,{\rm me-MR}}|$  to $|h'_{i,\rm me}|$, corresponding to the case when the ferromagnet has no intrinsic magnetostriction $b_1=b_2=0$). The results are shown for the same model parameters as in Fig.~\ref{fig:g}(a,b). The contribution of the magneto-rotation coupling is small, and therefore green contours are magnified 10 times. There is no magneto-rotational contribution to the in-plane component of the magnetoelastic field for R-SAW (c). }\label{fig:h}
\end{figure}

\begin{figure}
\includegraphics[width=\columnwidth]{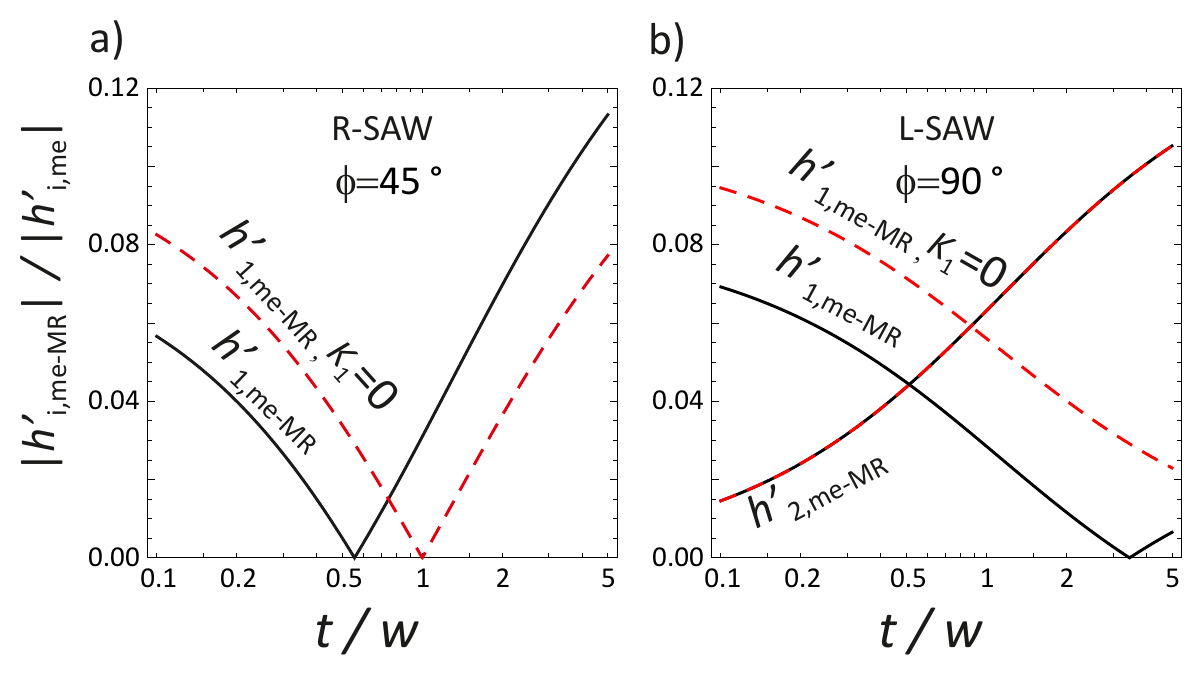}% Here is how to import EPS art
\caption{The relative contribution of the out-of-plane ($i=1$) and in-plane component ($i=2$) of dynamic magneto-rotation field $h'_{i,{\rm me-MR}}$, referred to the corresponding dynamic components of total magnetoelastic filed $h'_{i,{\rm me}}$ for (a) R-SAW (b) L-SAW. The values $|h'_{i,{\rm me-MR}}|/|h'_{i,{\rm me}}|$ are plotted depending on the thickness-to-width ratio ($t/w$) for two selected directions $\phi_{M}=45^{\rm o}$ and $\phi_{M}=90^{\rm o}$ where total magnetoelastic coupling is large for R-SAW and L-SAW, respectively. The solid-black and dashed-red lines denote the cases where the magnetocrystalline anisotropy is included and neglected ($K_1=0$), respectively. The results are shown for the same model parameters as in Fig.~\ref{fig:g}(c,d). %There is a magneto-rotation contribution to the in-plane component of the magnetoelastic field for R-SAW (a). 
}\label{fig:h_ratio}
\end{figure}

 %For a flat strip (small values of $t/w$), when cancelling the easy axis out-of-plane anisotropy $K_1=0$ the effective anisotropy increases. This effect is reflected in the increase of the magento-rotational coupling constants $|K_{xz}|$, $|K_{yx}|$
 
Once we neglect $K_1=K_{\rm s}/t$ (e.g. by the incense of the thickness $t$), we can focus on the shape anisotropy which only on the $t/w$ ratio and not on the absolute values of width $w$ and thickness $t$. Green contours in Fig.~\ref{fig:g}(c,d) are plotted for the same shape of the strip $t/w=0.1$ as in Fig.~\ref{fig:g}(a,b). %is larger due to the increase of $|K_{xz}|$ and $|K_{yx}|$.
 Let's discuss how the modification of the shape anisotropy, by the increase of the $t/w$ ratio affects the magneto-rotational coupling for R-SAW and L-SAW. This effect is illustrated by the orange, pink, and red contours in Fig.~\ref{fig:g}(c,d). For R-SAW, the magneto-rotation coupling is smaller than for L-SAW, changes non-monotonously, and is reduced to zero for $t/w= 1$. However, for L-SAW the magneto-rotation coupling strength changes differently. The magneto-rotational contribution to the energy density $|g_{\rm me}|$ grows with increasing $t/w$ ratio. It is worth noting that the lines in Fig.~\ref{fig:g} are not continuous for angles $\phi_{M}\approx 0$. This corresponds to the case when the external magnetic field $H_0$ (we used the value $H_0=0.5 M_{\rm s}$) cannot reorient the static magnetization near the direction of the hard axis $\phi_{M}= 0$ -- see Appendix A.

The analysis of the magnetoelastic energy density $|g_{\rm me}|$ obscures the role of the individual components of the magnetoelastic field. Fig.~\ref{fig:h} presents the angular dependence of the out-of-plane (Fig.~\ref{fig:h}(a,b)) and in-plane (Fig.~\ref{fig:h}(c,d)) components of total magnetoelastic field (black contours) and their magneto-rotational contribution (green contours). The results refer to the flat strip $t/w=0.1$ with out-of-plane magnetocrystalline anisotropy, that corresponds to the $|g_{\rm me}|$ in Fig.~\ref{fig:g}(a,b). It is easy to see that for R-SAW (Fig.~\ref{fig:h}(a,c)) the in-plane component of the magneto-rotational contribution to the magnetoelastic field $h'_{2,{\rm me-MR}}$ is zero, while for L-SAW (Fig.~\ref{fig:h}(b,d)) it is significantly reduced compared to the out-of-plane component $h'_{1,{\rm me-MR}}$. In the considered system (i.e., for a planar strip embedded in an elastic substrate), the magneto-rotation effects affect the magnetization dynamics mainly due to the out-of-plane component of the effective field.

Let's discuss more strictly the modification of the components of the magnetoelastic field due to magneto-rotation coupling. Fig.~\ref{fig:h_ratio} preset the dependence of the ratio of magneto-rotational contribution to the total magnetoelastic field of out-of-plane and in-plane components for R-SAW (Fig.~\ref{fig:h_ratio}(a)) and L-SAW (Fig.~\ref{fig:h_ratio}(b)). We selected the directions $\phi_{M}=45^{\rm o}$ and $\phi_{M}=90^{\rm o}$ around which one can expect the largest magnetoelastic coupling for R-SAW and L-SAW, respectively. 

For R-SAW, the in-plane component is zero ($h'_{2,{\rm me-MR}}=0$) and relative strength of the  out-of-plane  component $h'_{1,{\rm me-MR}}/h'_{1,{\rm me}}$ changes non-monotonously, reaching zero at $t/w \ne 1$ ($t/w=1$) for $K_{1} \ne 0$  ($K_{1}=0$). However, for L-SAW, the relative contribution $h'_{1,{\rm me-MR}}/h'_{1,{\rm me}}$ ($h'_{2,{\rm me-MR}}/h'_{2,{\rm me}}$) changes monotonously decreasing (increasing) with growing $t/w$ ratio. 

\section{Conclusions}

We studied the magneto-rotational coupling in a ferromagnetic strip. Our analysis demonstrated that all non-diagonal coefficients of the magneto-rotational coupling matrix are non-zero for this system and can be tailored by adjusting the shape anisotropy, which depends on the ratio of the strip’s thickness to its width. 

We investigated how the coupling between the fundamental mode of magnetization in a strip embedded near the surface of a non-magnetic material and surface acoustic waves of the Rayleigh or Love type depends on the direction of magnetization. The magneto-rotational field components, oriented perpendicular to the surface, play a dominant role (Fig.~\ref{fig:h}(a,b)). The angular characteristics of these fields are orthogonal for Rayleigh and Love waves. For a Rayleigh wave, the magneto-rotational coupling is strongest when the magnetization is aligned with the wave propagation direction (where conventional coupling is weakest): $\phi_{M}=0$. In contrast, for a Love wave, the magneto-rotational interaction is most pronounced when the magnetization is oriented perpendicular to the wave propagation direction: $\phi_{M}=90^{\rm o}$.

The magneto-rotational interaction is weaker compared to conventional magnetostriction. When the magnetic field is aligned in the direction that maximizes conventional magnetoelastic interaction, the magneto-rotational contribution  $|\mathbf{h}'_{{\rm me-MR}}|/|\mathbf{h}'_{{\rm me}}|$ constitutes only a few percent of the magnetoelastic field. For the system under study -- a CoFeB strip with a SAW propagating perpendicular to its axis -- this contribution is approximately 8\% when the thickness-to-width ratio $t/w$ approaches zero (where the strip resembles a layer)%\textcolor{red}{\st{ForRayleigh wave, the contribution decreases with a reduction in $t/w$ (Fig.~\ref{fig:h_ratio}).}}

For a Rayleigh wave, in the absence of magnetocrystalline anisotropy, the magneto-rotational contribution to the magnetoelastic field reaches a minimum at $t/w=1$. Beyond this point (for $t/w>1$), the contribution increases significantly (Fig.~\ref{fig:h_ratio}(a)). For Love wave, the contribution of the out-of-plane component decreases with a reduction in $t/w$, whereas the in-plane component increases (Fig.~\ref{fig:h_ratio}(b)).

%\begin{widetext}
%\end{widetext}

\begin{acknowledgments}
This work has received funding from National Science Centre Poland grants  UMO-2020/39/O/ST5/02110,\break UMO-2021/43/I/ST3/00550, and support from the Polish National Agency for Academic Exchange grant BPN/PRE/2022/1/00014/U/00001.  The authors would like to thank Dr. Piotr Graczyk for his comments and remarks.
\end{acknowledgments}

\section*{Conflict of Interest Statement}
There are no conflicts to declare.

\section*{Author contributions}
G. C. Data Curation, Formal Analysis, Funding Acquisition, Investigation, Methodology, Software, Visualization, Writing/Original Draft Preparation, Writing/Review \& Editing
J. W. K. Conceptualization, Data Curation, Formal Analysis, Funding Acquisition, Methodology, Investigation, Project Administration, Resources, Software, Supervision, Validation, Visualization, Writing/Original Draft Preparation, Writing/Review \& Editing

\section*{Data Availability Statement}
The data for the essential figures (Fig.~\ref{fig:K}, Fig.~\ref{fig:g}) can be accessed via the following URL https://doi.org/10.5281/zenodo.15471190. The remaining data is available from the corresponding author on reasonable request.

\renewcommand{\theequation}{{A}\arabic{equation}}
\setcounter{equation}{0}

\section*{Appendix A: Direction of equilibrium magnetization in ferromagnetic strip}\label{Appendix A}

\begin{figure}

\includegraphics[width=.75\columnwidth]{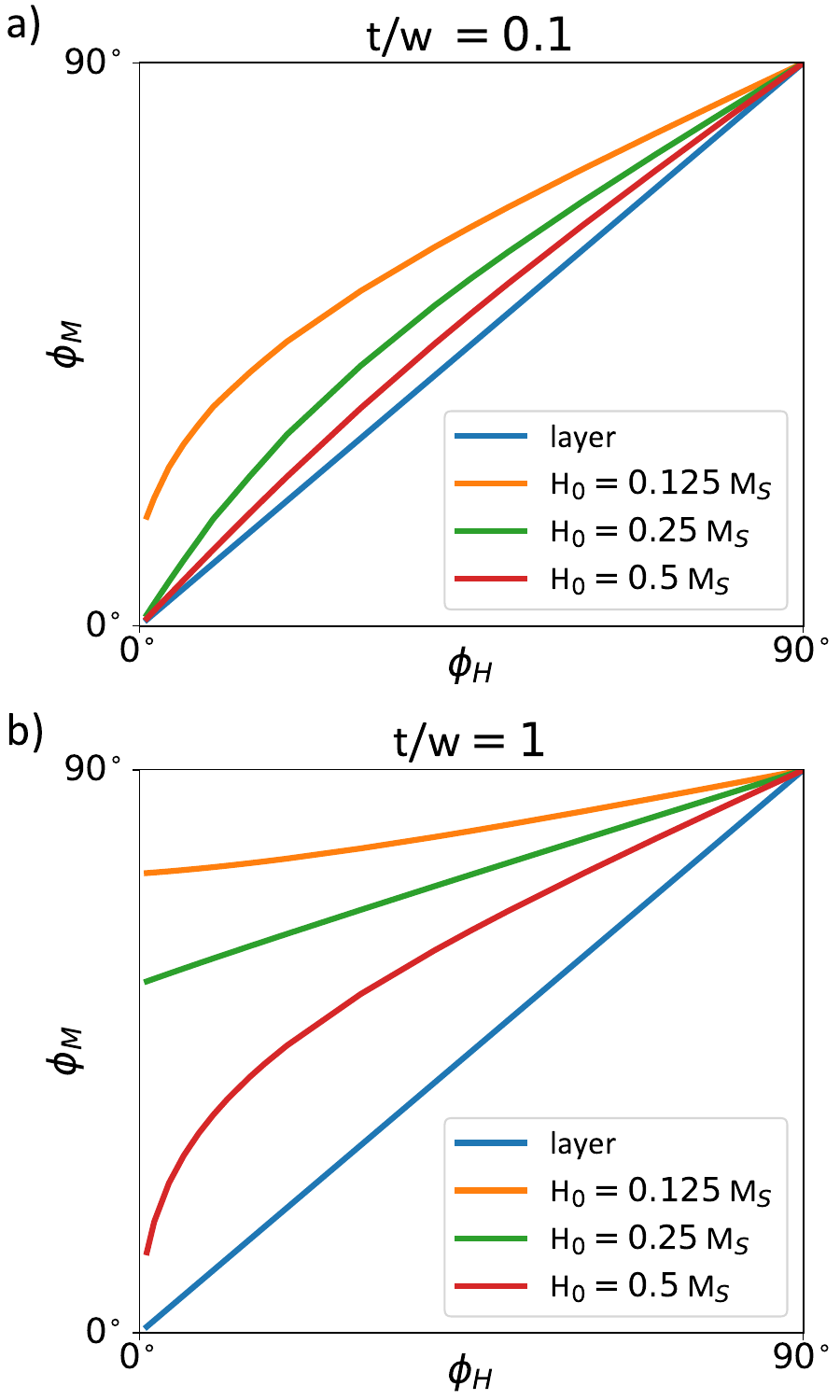}% Here is how to import EPS art
\caption{\label{fig:SI_angle_dependency} 
The angular dependence of the orientation of equilibrium magnetization ($\phi_{M}$) with respect to the direction
applied external magnetic field ($\phi_{H}$) for (a) flat: $t/w=0.1$ and (b) square $t/w=1$) strip, respectively. The dependences $\phi_M(\phi_H)$ are plotted for a few values of the external magnetic field $H_0$ equal to 0.125, 0.25, 0.5 of the saturation magnetization (M$_{\rm s}$).  
}
\end{figure}

In the absence of other sources of anisotropy and an external magnetic field, the shape anisotropy forces the magnetization to align along the axis of the strip. The application of an external magnetic field deflected from the strip axis, can change the direction of equilibrium magnetization. However, this change depends on the value of the external magnetic field, and only for very strong field the orientation of magnetization $\phi_M$ follows the applied field direction $\phi_H$ -- see Fig.~\ref{fig:SI_angle_dependency}~(a, b). 

\begin{comment}
\begin{figure}
\includegraphics[width=.75\columnwidth]{fig_1b_appendix_cal.pdf}% Here is how to import EPS art
\caption{\label{fig:profiles} 
a) The amplitude of the dynamic component of magnetization in the out-of-plane direction (fundamental mode), for angles of application of the external magnetic field $\phi=20^{\circ},45^{\circ},70^{\circ}$. For each angle, the ratio of the average value of the in-plane dynamic magnetization amplitude $<|m_{\parallel}|>$ and the average value of the out-of-plane dynamic magnetization amplitude $<|m_{\perp}|>$ expressing the average ellipticity was determined. b) The change in the fundamental mode frequency with the angle of magnetization ($\phi_{M}$), in the strip with $t/w$ ratio 0.1 (pink dashed line) and 1 (brown dashed line). The black line is the FMR frequency for the layer of pristine CoFeB. Results where obtained for $H_0=0.25~M_{\rm s}$.
}
\end{figure}

\end{comment}

\section*{Appendix B: Amplitudes of dynamic magnetization}\label{Appendix B}

To calculate the magnetoelastic energy density, it is necessary to know the amplitudes of the in-plane and out-of-plane components of magnetization. However, these amplitudes are not homogeneous for strips with a cross-section different from elliptical. Fortunately, the inhomogeneities in the profile of the fundamental mode are not large for the considered strip sizes with $t\times w< 25000$ nm$^2$, and we can assume that the precession of magnetization is approximately homogeneous across the strip. We used COMSOL Multiphysics (finite element method) to obtain the profiles of the fundamental mode and to calculate the averaged ellipticity of the fundamental modes. The obtained values are close to those for a strip with an elliptical cross-section, which can be obtained from the analytical formula\cite{gurevich_magnetization_2020}:
\begin{equation}\label{eq:ellipticity}
    \frac{m_2}{m_1}=i\sqrt{\frac{-N_{yy}\cos2\phi_M+H_0/M_{\rm s}\cos\phi_{H}}{N_{zz}\!-\!2K_1/(\mu_0 M_{\rm s}^2)\!-\!N_{yy}\sin^2\phi_M\!+\!H_0/M_{\rm s}\cos\phi_H}},
\end{equation}
where $N_{yy}$ and $N_{zz}=1-N_{yy}$ are elements of the demagnetization tensor. 

%\nocite{*}

\section*{References}

%\def\bibsection{\section*{\refname}} 
%\bibliographystyle{aipnum4-1}
%\bibliography{biblio}% Produces the bibliography via BibTeX.

%

\end{document}